\documentclass[twocolumn]{aastex631}
\usepackage{mathrsfs}
\usepackage{graphicx}
\usepackage{subfigure}
\usepackage{epstopdf}
\usepackage{amsmath}
\usepackage{amssymb}
\usepackage{hyperref}
\usepackage{color}
\usepackage{bbding}
\usepackage{fourier}

\begin{document}
\title{Evidence for Intermediate-Mass Black Holes From Microlensing Signatures in CHIME/FRB Catalog 2}

\author{Huan Zhou}
\affiliation{School of Physics and Optoelectronic Engineering, Yangtze University, Jingzhou, 434023, China}

\author{Zhengxiang Li}
\affiliation{School of Physics and Astronomy, Beijing Normal University, Beijing 100875, China}
\affiliation{Institute for Frontiers in Astronomy and Astrophysics, Beijing Normal University, Beijing
102206, China}
\email{zxli918@bnu.edu.cn}

\author{Cheng-Gang Shao}
\affiliation{School of Physics and Optoelectronic Engineering, Yangtze University, Jingzhou, 434023, China}
\email{cgshao@hust.edu.cn}

\author{Xi-Jing Wang}
\affiliation{School of Physics and Technology, Wuhan University, Wuhan 430072, China}

\author{Kai Liao}
\affiliation{School of Physics and Technology, Wuhan University, Wuhan 430072, China}

\author{He Gao}
\affiliation{School of Physics and Astronomy, Beijing Normal University, Beijing 100875, China}
\affiliation{Institute for Frontiers in Astronomy and Astrophysics, Beijing Normal University, Beijing
102206, China}

%\author{Bing Zhang}
%\affiliation{The Hong Kong Institute for Astronomy and Astrophysics, The University of Hong Kong, Pokfulam Road, Hong Kong, People’s Republic of China}
%\affiliation{Department of Physics, The University of Hong Kong, Pokfulam Road, Hong Kong, People’s Republic of China}
%\affiliation{The Nevada Center for Astrophysics and Department of Physics and Astronomy, University of Nevada, Las Vegas, NV, 89154, USA}

\author{Zong-Hong Zhu}
\affiliation{School of Physics and Astronomy, Beijing Normal University, Beijing 100875, China}
\affiliation{Institute for Frontiers in Astronomy and Astrophysics, Beijing Normal University, Beijing
102206, China}
\affiliation{School of Physics and Technology, Wuhan University, Wuhan 430072, China}
%\email{zhuzh@whu.edu.cn}

\begin{abstract}
Intermediate-mass black holes (IMBHs) are the missing link in the cosmic hierarchy of black holes, bridging the gap between stellar-mass black holes and supermassive ones. They also serve as unique laboratories for testing strong-field gravity and are prime targets for future multi-messenger observations. However, IMBHs are a population that has remained notoriously difficult to detect. The microlensing effect of fast radio bursts (FRBs) can serve as a clean and powerful method to probe IMBHs. In this work, we develop a pipeline to search for microlensed FRBs based on their dynamic spectra and apply it to the CHIME/FRB Catalog 2. Two microlensing signatures have been identified in two separate sources, i.e. FRB~20190131D and FRB~20211115A. The inferred lens masses for these two signatures are $\sim[280-467]~M_{\odot}$ and $\sim[539-609]~M_{\odot}$, respectively. Here we interpret them as evidence for IMBHs. If there are no intervening structures-such as galaxies or clusters-along the line of sights for these two sources, the two identified IMBHs might be isolated and of primordial origins. In that case, we obtain primordial black holes (PBHs) within these two mass ranges would constitute $\sim4\%$ of dark matter. Moreover, if these two candidates are not genuine lensing signatures, the abundance of intermediate-mass PBHs with masses $>300,M_{\odot}$ is constrained to be $\sim13\%$ at $95\%$ confidence level. Therefore, more comprehensive observational information for FRBs, together with a deeper understanding of whether the intrinsic emission mechanisms of FRBs can produce lensing-like signals, will be crucial for establishing this effect as a powerful tool for probing (primordial) IMBHs.
\end{abstract}

\section{Introduction}
Intermediate-mass black holes (IMBHs), defined as black holes with masses between $10^2$ and $10^5~M_{\odot}$, occupy a critical yet poorly explored regime in the cosmic black hole mass function, bridging the gap between stellar-mass black holes ($\lesssim 10^2\,M_{\odot}$) and supermassive black holes ($\gtrsim 10^6\,M_{\odot}$). Candidate IMBHs have been reported through multiple observational channels, each with distinct strengths and limitations: (i) radiative accretion signatures from ultraluminous X-ray sources~\citep{Tremou:2018rvq,Lin:2018dev}, which can also be explained by super-Eddington accretion onto stellar-mass black holes; (ii) stellar kinematics from integral field spectroscopy~\citep{2022ApJ92448P,2024Natur631285H,2025NSRev12E347H}, limited by spatial resolution requirements; (iii) gravitational-wave detections of binary black hole mergers in the IMBH regime, most notably GW190521 and GW231123, which provide the most direct evidence by encoding component masses without astrophysical modeling degeneracies~\citep{LIGOScientific:2020iuh,LIGOScientific:2025rsn}; and (iv) microlensing of extragalactic transients~\citep{Paynter:2021wmb,LIGOScientific:2025cwb,Xiong:2025gtw}, whose interpretation depends on the assumed lens model. Despite these efforts, robust confirmation of a substantial IMBH population remains challenging, as most candidates suffer from degeneracies with other astrophysical systems, and the IMBH mass function remains observationally unconstrained.

From a theoretical perspective, one of the leading formation channels for IMBHs is dynamical assembly in dense star clusters, which circumvents the mass limits of stellar collapse. Three key mechanisms have been identified: (i) runaway stellar collisions in dense cluster cores producing a very massive star ($\gtrsim 10^3\,M_{\odot}$) that collapses directly into an IMBH~\citep{PortegiesZwart:2004ggg}, most efficient at low metallicities ($Z \lesssim 0.1\,Z_{\odot}$) and high cluster densities ($\gtrsim 10^5\,M_{\odot}\,{\rm pc}^{-3}$); (ii) hierarchical mergers of stellar-mass black holes that gradually build up an IMBH~\citep{ArcaSedda:2023mlv,Paiella:2025tia}, with efficiency depending on cluster retention fraction and binary availability; and (iii) tidal disruption events that contribute to black hole growth over time~\citep{Rizzuto:2022fdp}, though likely subdominant. The relative importance of these channels remains an active area of investigation, with implications for the predicted IMBH mass function and its redshift evolution.

In addition to the above-mentioned formation channels, primordial black holes (PBHs), formed in the early Universe via the collapse of primordial density fluctuations~\citep{Hawking:1971ei,Carr:1974nx,Carr:1975qj}, are also a possible origin of IMBHs. Meanwhile, PBHs are a well-motivated dark matter candidate whose mass range spans over 50 orders of magnitude---from the Planck scale ($\sim 10^{-5}\,{\rm g}$) to supermassive black hole masses ($\gtrsim 10^{9}\,M_{\odot}$). Recent multi-wavelength and multi-messenger advances have positioned PBHs as a compelling theoretical alternative for interpreting various astrophysical phenomena, including the origin of IMBHs. In GW astronomy, the LIGO-Virgo-KAGRA Collaboration has detected several BBH merger events with component masses in the IMBH regime~\citep{LIGOScientific:2020iuh,LIGOScientific:2025rsn}, most notably GW190521 and GW231123, whose masses and spins are difficult to explain via conventional stellar evolution but can be reproduced by the PBH merger channel when cosmological accretion is taken into account~\citep{DeLuca:2025fln}. Similarly, recent JWST observations have revealed a population of quasars at $z \gtrsim 6$~\citep[e.g.,][]{Maiolino:2023bpi,Bogdan:2023ilu,Natarajan:2023rxq}, with Abell2744-QSO1---hosting an SMBH of $\sim 5\times 10^7\,M_{\odot}$ with an extreme black hole-to-stellar mass ratio and low metallicity---posing severe challenges to conventional SMBH formation models~\citep{Maiolino:2025tih,Juodzbalis2025}. These anomalies motivate alternative seeding scenarios, including PBHs, which could potentially account for such systems under certain accretion and evolution conditions~\citep{Dayal:2025aiv,Zhang:2025oyl,DeLuca:2025nao}. Extensive observational searches have placed constraints on the PBH abundance $f_{\mathrm{PBH}} \equiv \Omega_{\mathrm{PBH}} / \Omega_{\mathrm{DM}}$ across a wide range of mass windows~\citep{Sasaki2018,Green:2020jor,Carr:2020gox,Carr:2021bzv,2026NCimRtmp4C}; for masses $\gtrsim 100\,M_{\odot}$, current limits from microlensing and GW observations constrain $f_{\mathrm{PBH}} \lesssim 0.1{-}1$, depending on the mass range and assumed mass function. Among various methods, gravitational lensing provides one of the most direct probes of PBH abundance, sensitive to masses from planetary scales ($\sim 10^{-7}\,M_{\odot}$) to SMBHs, and can be categorized into four types: (1) microlensing-induced flux variations of stars~\citep{Niikura:2017zjd,Mroz:2024mse}; (2) time-delayed multi-peak structures in transients such as FRBs and GRBs~\citep{Munoz:2016tmg,Liao:2020wae}; (3) waveform distortions in coherent transients due to lensing interference~\citep{Jung:2017flg,CHIMEFRB:2022xzl}; and (4) angularly separated multiple images of persistent sources~\citep{Wilkinson:2001vv,Zhou:2021tvp}.

Fast radio bursts (FRBs) are millisecond-duration radio pulses of cosmological origin, with isotropic-equivalent energies spanning $\sim 10^{36}{-}10^{41}\,{\rm erg}$~\citep{Lorimer:2007qn}. Only $\sim 2\%{-}3\%$ repeat~\citep{CHIMEFRB:2023myn,TheCHIMEFRB:2026nji}, likely a lower bound; the intrinsic fraction may exceed $50\%$~\citep{Yamasaki:2023dlb}. The Galactic FRB associated with SGR 1935+2154 confirmed magnetar engines as one progenitor channel~\citep{Zhang:2020qgp,CHIMEFRB:2020abu,Bochenek:2020zxn,Lin:2020mpw}, though the broader radiation mechanism remains debated. Despite these uncertainties, FRBs' clean temporal structure, cosmological distances, and high event rate ($\sim 10^{3}{-}10^{4}\,{\rm sky^{-1}\,day^{-1}}$~\citep{Cordes:2019cmq,Petroff:2019tty}) make them promising probes for cosmology~\citep{Deng:2013aga,Liu:2022bmn,Wei:2015hwd,Wu:2016brq,Li:2017mek} and compact objects.~\citet{Munoz:2016tmg} proposed using FRB microlensing to probe IMBHs, and subsequent searches have been conducted~\citep{Liao:2020wae,Zhou:2021ndx,Krochek:2021opq,Chang2025,Xiong:2025gtw}, identifying FRB~20190308C and FRB~20190320B as possible candidates---though both remain debated. As next-generation facilities (CHIME, DSA-2000, SKA) expand the FRB sample, FRBs will become increasingly powerful for constraining PBHs~\citep{Munoz:2016tmg,Laha:2018zav,Oguri:2022fir,Connor:2022bwl}. 

In this paper, we refine the microlensing search methodology and apply it to CHIME/FRB Catalog 2~\citep{TheCHIMEFRB:2026nji} to search for lensing signals and update their implications on IMBHs and abundace of PBHs.

\section{FRB Microlensing Candidates}\label{sec2}
In this section, we briefly introduce the current status of CHIME/FRB observations, present the method for searching and identifying microlensing signatures, and report the search results along with an analysis of the candidate properties.

\subsection{FRB Observations in CHIME/FRB Catalog 2}\label{sec2-1}
The rapid increase in the number of verified FRBs is currently being driven by the operation of several wide-field radio telescopes, such as Canadian Hydrogen Intensity
Mapping Experiment (CHIME), Five-hundred-meter Aperture Spherical radio Telescope (FAST), the Australian Square Kilometre Array Pathfinder (ASKAP), and the Deep Synoptic Array (DSA). As a prominent example, the CHIME/FRB Collaboration initially released its first catalog containing 535 FRBs detected in less than one year (from July 25, 2018, to July 1, 2019)~\citep{CHIMEFRB:2021srp}. Subsequently, a second catalog (CHIME/FRB Catalog 2) was published, comprising 4539 FRBs observed with the CHIME telescope between July 25, 2018 and September 15, 2023~\citep{TheCHIMEFRB:2026nji}. The second catalog includes all FRBs from the first catalog, with every event reprocessed using a uniform and improved analysis framework. The 4539 bursts originate from 3641 unique sources: 3558 non‑repeating sources, 83 known repeating sources contributing a total of 981 bursts~\footnote{The basic information for all FRBs in the CHIME/FRB Catalog~2 is available at~\url{https://www.chime-frb.ca/catalog2}.}.

\begin{figure}[ht!]
    \centering
    \includegraphics[width=0.45\textwidth, height=0.36\textwidth]{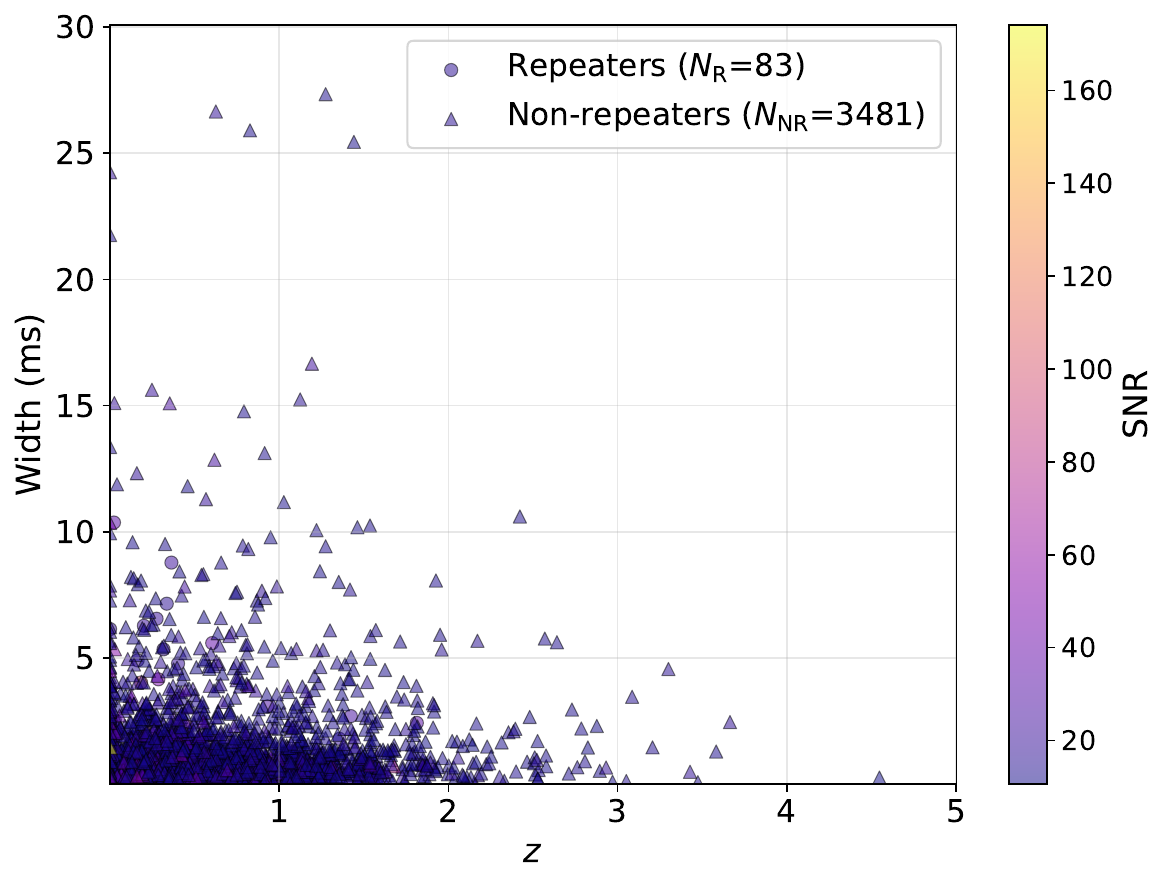}
     \caption{Two-dimensional distribution of inferred redshifts, width, and SNR for 3564 FRBs in CHIME/FRB catalog 2.}\label{fig1}
\end{figure}

For a detected FRB, the dispersion measure (DM) is a key observational property. Theoretically, DM is the integral of the electron density along the radio pulse's path. Observationally, it is derived from the arrival time difference between two photons at different frequencies. The high DMs of the first few poorly localized bursts suggested a cosmological origin for FRBs~\citep{Lorimer:2007qn,Thornton:2013iua}. This was later confirmed when the repeating FRB 20121102A was localized to a nearby dwarf galaxy~\citep{Chatterjee:2017dqg,Tendulkar:2017vuq,Marcote:2017wan}. Therefore, the distance and the redshift can be roughly derived from the observed DM of a detected FRB, which is usually decomposed into four components: the Milky Way (${\rm D_{MW}}$), the intergalactic medium (${\rm DM_{IGM}}$), and the host galaxy including the source local environment (${\rm DM_{host}}$ and ${\rm DM_{src}}$), respectively
\begin{equation}\label{eq2-1}
{\rm DM}={\rm DM_{MW}}+{\rm DM_{IGM}}+\frac{\rm DM_{\rm host}+DM_{\rm src}}{1+z}.
\end{equation}
Here, we conservatively adopt a maximum value of ${\rm DM_{host}}+{\rm DM_{src}} = 200;\mathrm{pc,cm^{-3}}$, which corresponds to the minimum inferred redshift for all host galaxies. The ${\rm DM_{IGM}}-z$ relation, given by~\citet{Deng:2013aga} and approximated as ${\rm DM_{IGM}}\approx 855z;\mathrm{pc,cm^{-3}}$~\citep{Zhang:2018csb}, takes into account the helium ionization history and assumes an intergalactic medium (IGM) baryon fraction $f_{\rm IGM}=0.83$. This relation is statistically supported by the five localized FRBs available at the time~\citep{Li:2020qei}. The key observational features of FRBs and the inferred redshifts for the lensing scales, based on the currently public FRBs from the CHIME/FRB Catalog 2, are shown in Fig~\ref{fig1}. It is worth noting that among the 3558 non‑repeating sources, 77 sources lack width and SNR information and are therefore excluded from further analysis owing to observational issues, such as the inability to fit burst morphology from heavy Radio Frequency Interference (RFI) or background confusion, events on low‑sensitivity days (flagged by pulsar‑based noise monitoring), and events during periods when the real‑time L2/L3 classification was non‑functional.

\begin{figure*}
   \centering
   \includegraphics[width=0.8\textwidth, height=0.45\textwidth]{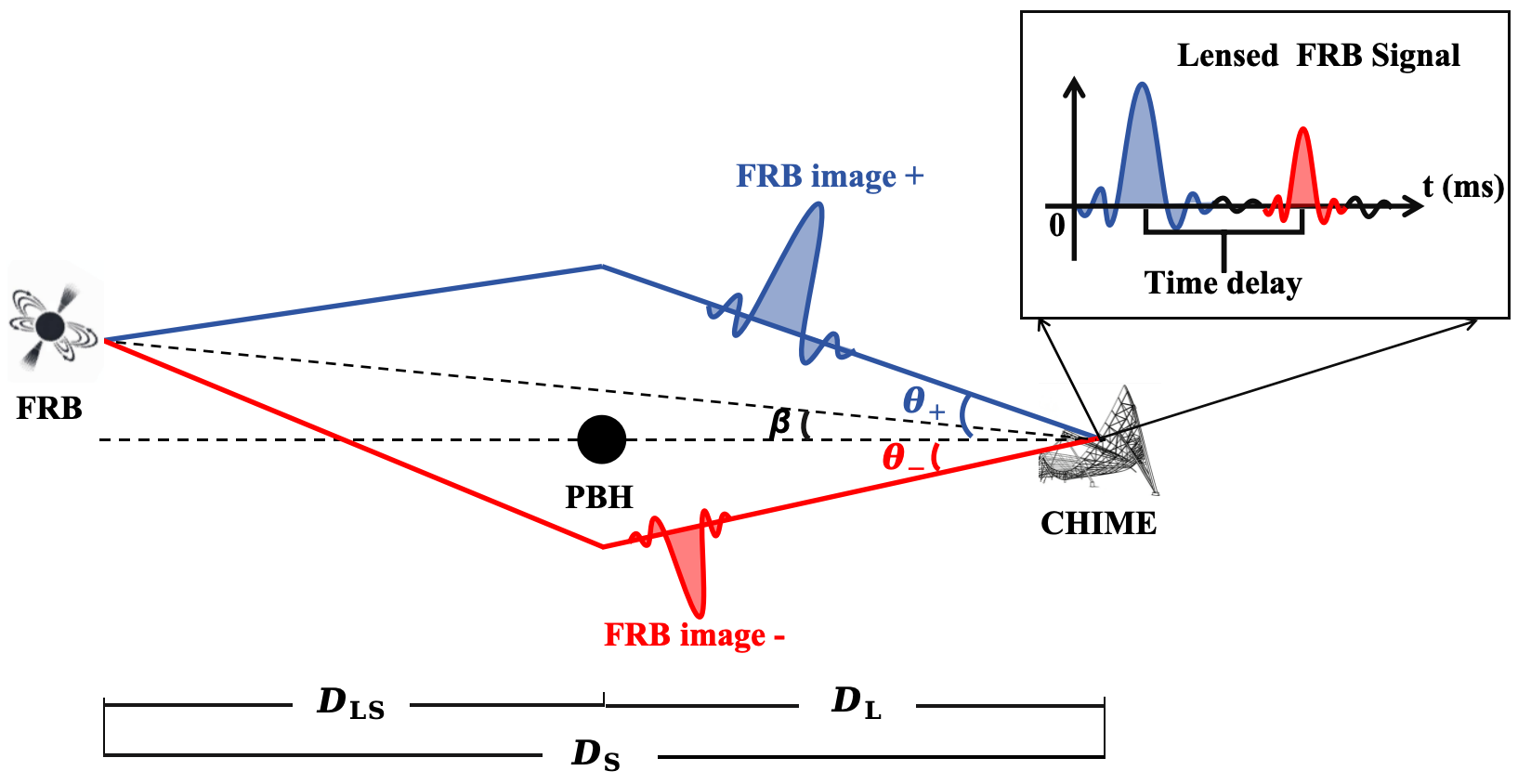}
     \caption{Schematic diagram of the FRB being lensed considering PBH as point-mass lens.}\label{fig2}
\end{figure*}
\subsection{Searching for Microlensing Signatures}\label{sec2-2}
Previous studies have established the foundation for searching microlensed FRBs by exploiting characteristic signatures such as similar light-curve morphology and the absence of significant frequency drift~\citep{Liao:2020wae, Zhou:2021ndx}, and have further applied correlation methods to search for such candidate events~\citep{Zhou:2021ndx, Krochek:2021opq, Chang2025, Xiong:2025gtw}. Building on these efforts, we develop a complete screening pipeline that systematically incorporates noise treatment and selection-bias mitigation, while accounting for the intrinsic features of microlensed FRBs. The pipeline consists of four sequential tests described in the following subsections: Autocorrelation Test, Peak‑SNR Test, Frequency‑Drift Test, and Hardness Test. This approach is designed for the current CHIME/FRB Catalog 2 and can be readily extended to larger future FRB samples.

\subsubsection{Autocorrelation Test}
\begin{figure*}
    \centering
    \includegraphics[width=0.45\textwidth, height=0.36\textwidth]
    {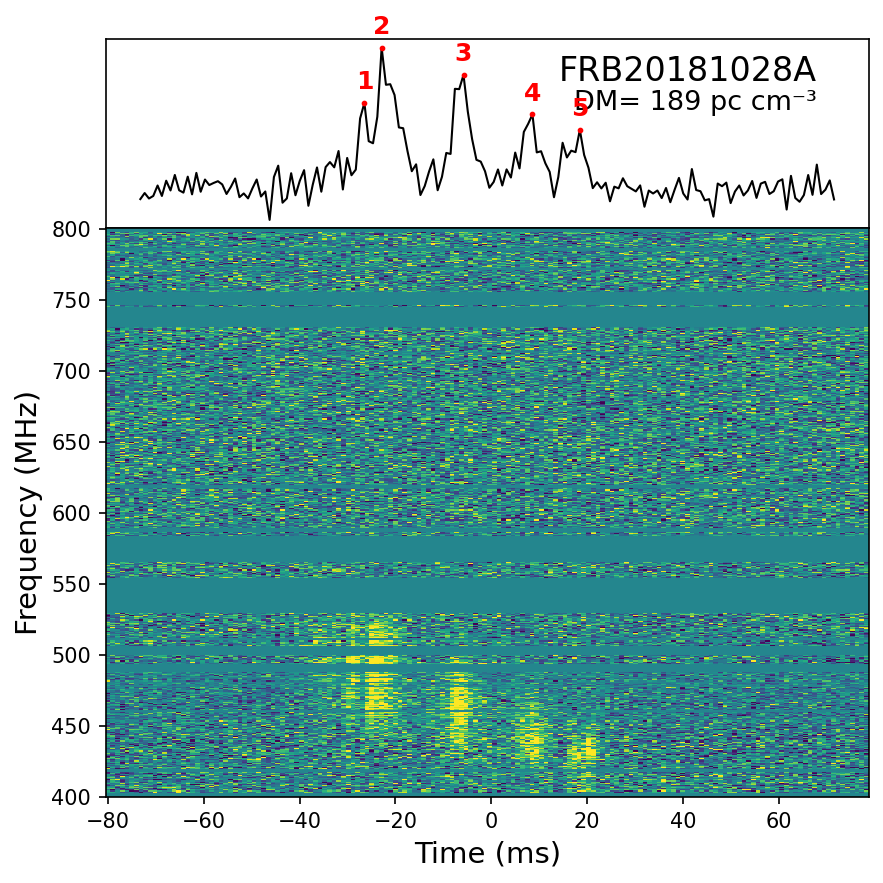}
    \includegraphics[width=0.45\textwidth, height=0.36\textwidth]
    {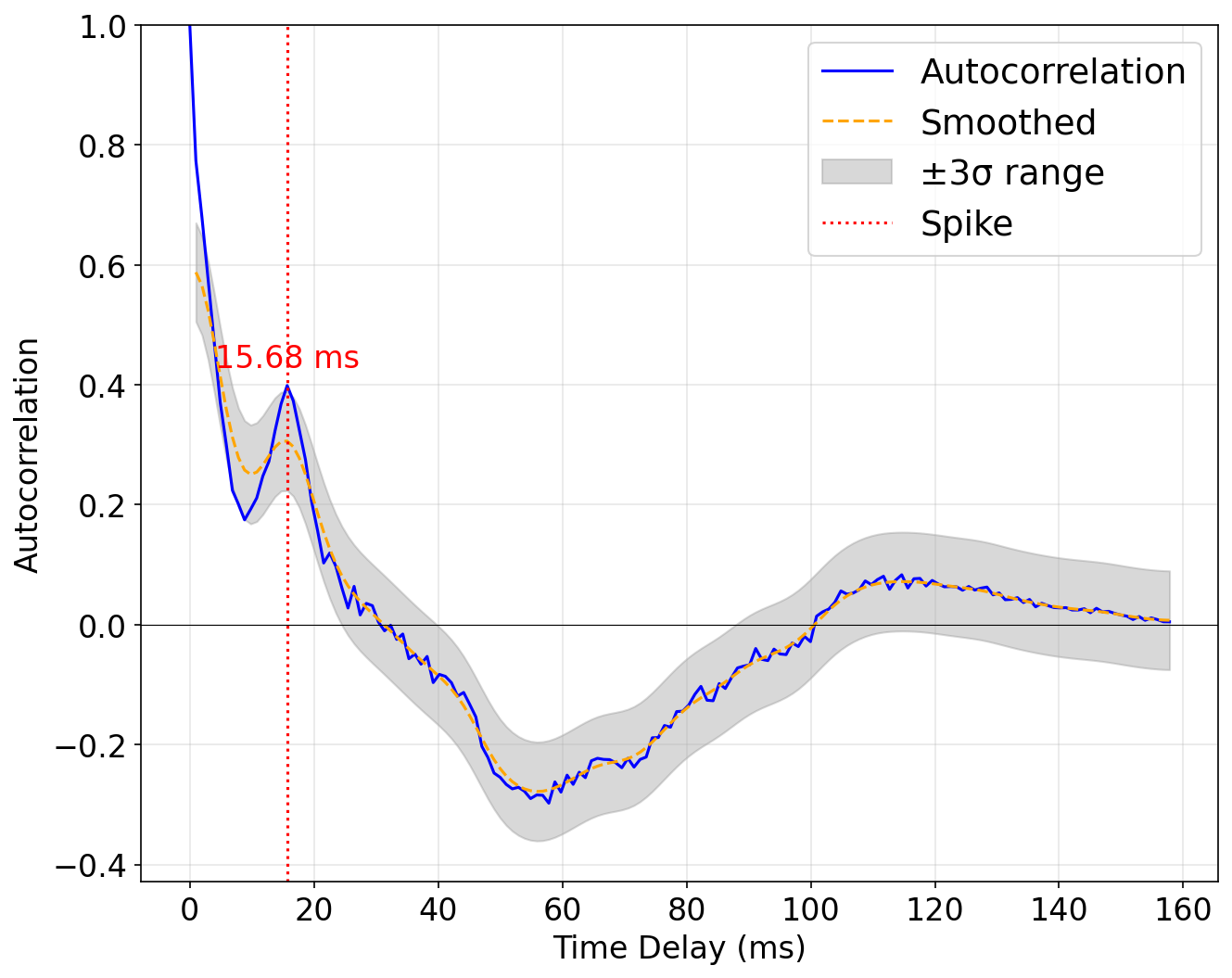}
    \includegraphics[width=0.45\textwidth, height=0.36\textwidth]{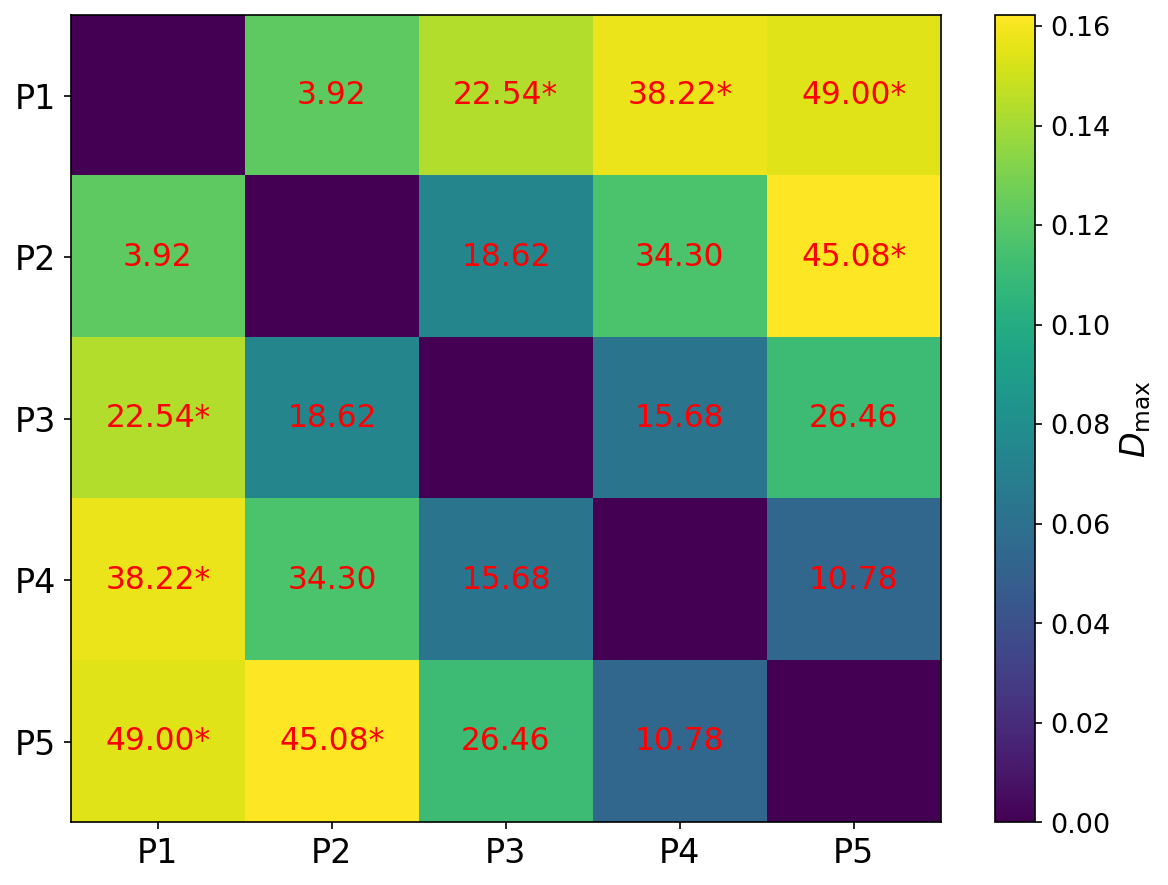}
    \includegraphics[width=0.45\textwidth, height=0.36\textwidth]{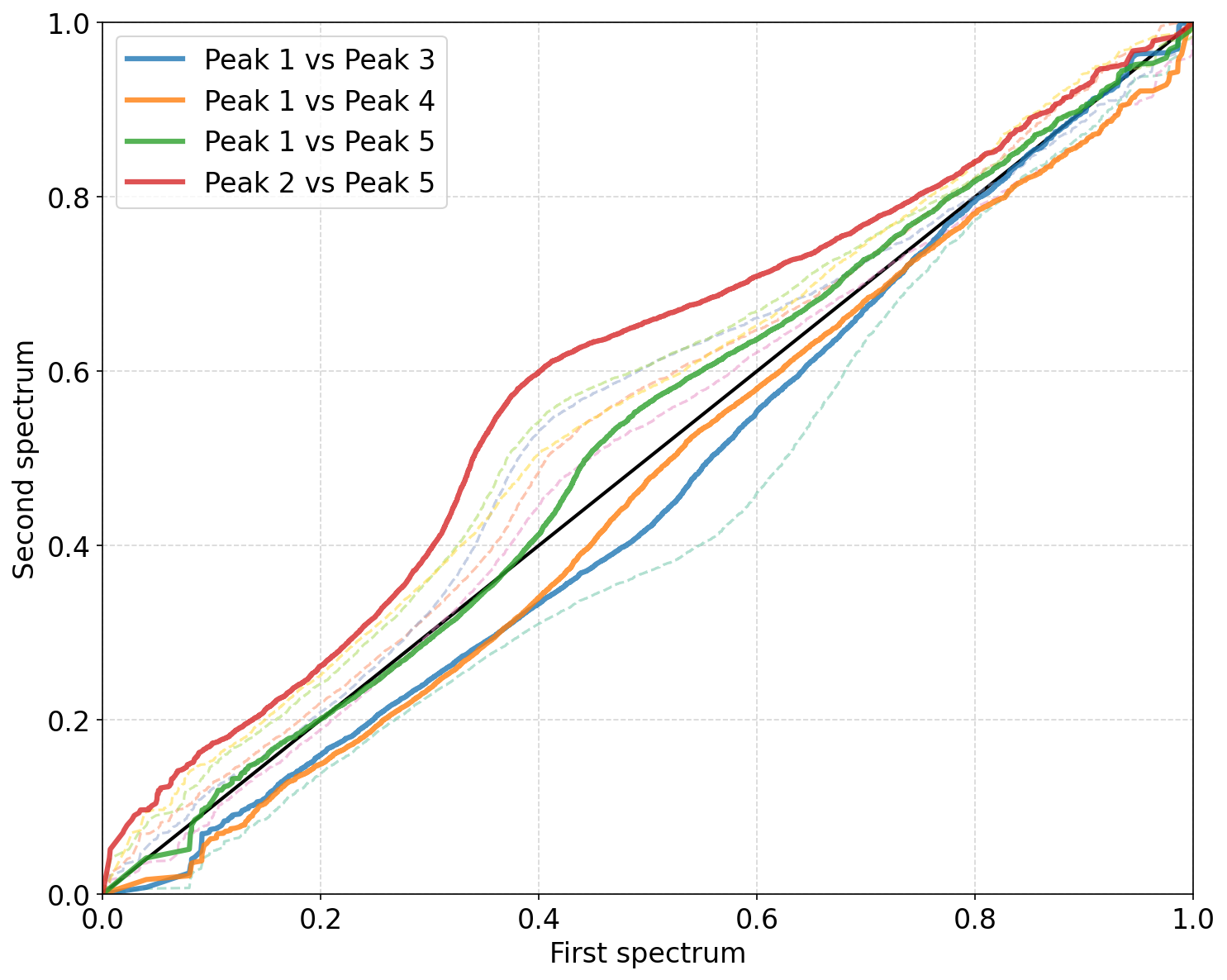}
     \caption{{\bf Upper Left:} The dynamic spectrum of FRB~20181028A without the RFI. 
     {\bf Upper Right:} The blue line shows the autocorrelation of the light curve. The orange dashed line represents a Gaussian-smoothed version of the autocorrelation. The gray dash-dotted lines indicate the $3\sigma$. The red dashed line marks a location with a strong autocorrelation exceeding the $3\sigma$ threshold, corresponding to a time delay of $\Delta t = 15.68~\mathrm{ms}$.
     {\bf Lower Left:} As indicated by the peaks marked in the dynamic spectrum, the heatmap shows the differences in the CDFs between peaks as a function of frequency separation. The color scale represents the maximum distance $D_{\rm max}$ from the K-S test, while the numerical values indicate the separation between peaks. A red star is assigned to any pair of peaks where $D_{\rm max} > 0.1$ and exceeds the $3\sigma$ noise level, marking the presence of a severe frequency drift between peak $i$ and peak $j$.
     {\bf Lower Right:} For comparison with the heatmap, the Q–Q plot of the frequency spectral distribution between peaks is presented, where the highlighted solid lines indicate the peak pairs exhibiting significant frequency drift just as the peak pairs marked in the lower‑left heatmap.
     }\label{fig3}
\end{figure*}

Among FRBs, the most likely lensing signals are those with clear multi‑peak structures (sub-burst); the CHIME/FRB Catalog 2 identified 340 FRBs exhibiting such morphologies~\citep{TheCHIMEFRB:2026nji}. For intrinsically short-duration FRBs (with a duration of $1~\rm ms$), a microlensed burst would appear as two distinct peaks. Their time delay can be clearly read from the light curve, as long as the flux ratio is not too small. The light curve of a microlensed FRB will have an echo superimposed on it. To detect this from these FRBs with mulit-peaks, we define the normalized auto-correlation function (ACF) of light curve 
\begin{equation}\label{eq2-2}
C(\delta t)={\sum\limits_{t}}\frac{\widetilde{I}(t)\widetilde{I}(t-\delta  t)}{N_{\delta t}\sigma_{\rm I}^2},
\end{equation}
where $\widetilde{I}(t)=I(t)-\mu_{I}$ is  intensities of the light curve, where $\sigma_{\rm I}$ is the standard deviation for the light curve $I(t)$, $\delta t$ is relative displacement for autocorrelation, and $N_{\delta t}$ is the bin number of the light curve. The effectiveness of the autocorrelation analysis on lensed data can be shown as follows. Let $I_{\rm int}(t)$ be the intrinsic light curve of the FRB event and $C(\delta t)$ its autocorrelation. In a specific point-mass lensing configuration as shown in Fig~\ref{fig2}, we assign a time delay $\Delta t$ and a flux ratio $R_{\rm f}$ to the second image, yielding a lensed light curve
\begin{equation}\label{eq2-3}
I_{\rm lensed}(t,\Delta t, R_{\rm f})\propto\frac{R_{\rm f}}{R_{\rm f}+1}I_{\rm int}(t+\Delta t)+\frac{1}{R_{\rm f}+1}I_{\rm int}(t).
\end{equation}
Using this formula together with Eq.~(\ref{eq2-2}), the ACF of the lensed signal exhibits spikes at $\delta t = -\Delta t$, $0$, $+\Delta t$ with an amplitude ratio ${R_{\rm f}}/{(R_{\rm f}^2+1)}\;:\;1\;:\;{R_{\rm f}}/{(R_{\rm f}^2+1)}$. To distinguish lensing-induced spikes from noise, we define the sigma parameter for ACF of light curve~\citep{Ji:2018rvg,Paynter:2021wmb,Chang2025,Xiong:2025gtw}
\begin{equation}\label{eq2-4}
\sigma_{\delta t}=\sqrt{\frac{1}{N_{\delta t}}{\sum\limits_{\delta t\in \Delta t}\bigg(C(\delta t)-G(\delta t)\bigg)^2}},
\end{equation}
where $C(\delta t)$ is the ACF to be examined, which may correspond to either a lensed or an unlensed signal, $G(\delta t)$ is Gaussian-smoothed version (standard deviation of the Gaussian kernel $\sigma=3)$ ) of $C(\delta t)$~\footnote{We have verified that the Gaussian smoothing parameters used in the selection process have little effect on the resulting FRB candidates, and the two potential events (FRB~20190131D and FRB~20211115A) remain unaffected. While different fitting methods (e.g., Savitzky–Golay filtering of the ACF~\citep{Paynter:2021wmb,Chang2025,Xiong:2025gtw}) may introduce discrepancies, they do not exclude high‑SNR lensed signals. Independent tests employing the Savitzky–Golay filter corroborate the persistence of above two candidates.}, and $\Delta t\in[0,N_{\delta t}\Delta t_{\rm min}]$ ($\Delta t_{\rm min}$ represents the time resolution of CHIME, which is approximately $0.98~\rm ms$) denotes the time grid over which the calculation is performed. The parameter $\sigma$ quantifies the overall spikiness of $C(\delta t)$ relative to the smooth template $G(\delta t)$. If we identify a $\delta t$ for which $|C(\delta t) - G(\delta t)|$ exceeds a certain threshold i.e., $3\sigma$, we can claim evidence for a spike induced by lensing. In Fig.~\ref{fig3} we present the results of this analysis for FRB~20181028A, revealing a spike at $15.68~\rm ms$ that most likely corresponds to the time delay between the third and fourth peaks in the dynamic spectrum~\footnote{The dynamic spectra of all FRBs were obtained from the CHIME/FRB Catalog 2 Public Data, available at the Canadian Astronomy Data Centre: \url{https://www.canfar.net/storage/list/AstroDataCitationDOI/CISTI.CANFAR/25.0066/data}.}.

\subsubsection{Peak-SNR Test}
In addition to the screening criteria described above, for point-mass lenses such as PBHs illustrated in Fig.~\ref{fig2}, the lensed double peaks must satisfy the condition that the main peak arrives before the secondary peak. Therefore, we define the peak SNR of a peak in the dynamic spectrum as
\begin{equation}\label{eq2-5}
\text{PSNR} = \frac{I_{\rm peaks}(t)-\bar{I}_{\rm noise}}{\sigma_{\rm noise}},
\end{equation}
where $I_{\rm peaks}(t)$ represents the peak intensity, $\bar{I}_{\rm noise}$ is the mean noise intensity, and $\sigma_{\rm noise}$ denotes the standard deviation of the noise. The point-mass lens model requires that ${\rm PSNR}_{i}>{\rm PSNR}_{j}$ for $i<j$. Similarly, taking FRB~20181028A in Fig.~\ref{fig3} as an example, peak~1 is significantly weaker than the other peaks, so it cannot form a point-mass lensed signal pair with any of the other peaks.

\subsubsection{Frequency-Drift Test}
Nevertheless, the presence of a spike in the ACF of the light curve does not guarantee a lensing origin, two peaks with severe frequency drift are not indicative of gravitational lensing effect. Here we propose a Kolmogorov–Smirnov (K-S) test to verify whether severe frequency drift exists between the two peaks
\begin{equation}\label{eq2-6}
D_{\rm max}=\sup_{f}[{\rm CDF}(\bar{I}_{i}(f)), {\rm CDF}(\bar{I}_{j}(f))]~(i\neq j),
\end{equation}
where $\bar{I}_{i}(f)$ represents the normalized intensity profile of the $i$th peak in the frequency domain, which we treat as a random distribution in frequency, and CDF denotes its corresponding cumulative distribution function. At a significance level of $\alpha = 0.05$, given that the two peaks have equal sizes (we rebinned the 32 frequency channels of the CHIME data, resulting in $n_f = 512$ frequency channels for the subsequent analysis), the critical value $D_{\rm crit}$ is approximately $D_{\rm crit}=1.36 \times \sqrt{2/n_f}\approx0.1$. To further assess the resistance of $D_{\rm max}$ to outlier noise, we conducted a bootstrap analysis of the noise using $\mathcal{O}(10^3)$ iterations. Finally, a pair of peaks is characterized as exhibiting significant frequency drift if $D_{\rm max}$ satisfies both of the following conditions
\begin{equation}\label{eq2-7}
\begin{cases}
D_{\rm max} > D_{\rm crit}, \\
D_{\rm max} > D_{\rm n,upp},
\end{cases}
\end{equation}
where $D_{\rm n,upp}$ denotes the $3\sigma$ upper bound of the $D_{\rm n,max}$ statistic obtained from K-S tests on the noise realizations. As illustrated in Fig.~\ref{fig3} for FRB~20181028A, significant frequency drift is present between peak~1 and peaks~3–5, while the peak pair (peak~3-4) corresponding to the distinct ACF spike exhibits no significant drift. This conclusion is further supported by the Q‑Q plots, which reveal different patterns of drift among the drifting pairs — for instance, some show clear divergence at the head and tail quantiles, whereas others display an S‑shaped deviation around the diagonal.

\begin{table*}[!t]
\centering
\renewcommand{\arraystretch}{1.5}
\setlength{\tabcolsep}{6pt}
{\begin{tabular}{c|c|c|c|c|c|c|c|c|c|c}
\hline
FRB Name & SNR & \(\Delta t\) & $f$ range & \multicolumn{2}{c|}{HR\(_{\text{ML}}\)} & \multicolumn{2}{c|}{HR\(_{\text{HM}}\)} & $R_{\rm f}$ & $M_{{\rm L},z}$ & Rank \\
\cline{5-8}
& & (ms) & (MHz) & Episode 1 & Episode 2 & Episode 1 & Episode 2 &  & (\(M_{\odot}\)) & $S_{\rm Lens}$ \\
\hline
FRB~20190131D & 51.0 & 8.82  & [400.2, 800.2] & {0.99 ± 0.09} & {1.16 ± 0.26} & {0.63 ± 0.07} & {0.57 ± 0.15} & 2.57 & 466.50 & 2+4=6 \\
FRB~20190915E & 51.3 & 2.94  & [521.0, 800.2] & {3.51 ± 0.49} & {2.93 ± 0.53} & \underline{0.52 ± 0.05} & \underline{0.65 ± 0.07} & {\XSolidBrush} & {\XSolidBrush} & {\XSolidBrush} \\
FRB~20200603B & 57.4 & 9.80 & [400.2, 582.2] & \underline{0.60 ± 0.05} & \underline{0.44 ± 0.05} & {0.35 ± 0.07} & {0.35 ± 0.10} & {\XSolidBrush} & {\XSolidBrush} & {\XSolidBrush}\\
FRB~20210117D  & 29.9 & 2.94 & [414.5, 624.5] & \underline{1.72 ± 0.20} & \underline{0.78 ± 0.18} & {0.35 ± 0.07} & {0.53 ± 0.21} & {\XSolidBrush} & {\XSolidBrush} & {\XSolidBrush} \\
FRB~20210130C  & 37.7 & 2.94 & [503.1, 800.2] & \underline{2.31 ± 0.48} & \underline{1.14 ± 0.16} & {0.94 ± 0.14} & {0.77 ± 0.13} & {\XSolidBrush}& {\XSolidBrush} & {\XSolidBrush}\\
FRB~20211115A & 27.1 & 6.86  & [400.2, 565.5] & {1.47 ± 0.17} & {1.24 ± 0.24} & {0.37 ± 0.07} & {0.38 ± 0.12} & 1.77 & 609.45 & 2+1=3 \\
FRB~20220225C & 35.0 & 8.82  & [400.2, 525.5] & \underline{0.80 ± 0.10} & \underline{1.29 ± 0.21} & {0.48 ± 0.10} & {0.63 ± 0.11} & {\XSolidBrush} & {\XSolidBrush} & {\XSolidBrush} \\
FRB~20220424C & 36.2 & 6.86 & [400.2, 800.2] & {1.27 ± 0.14} & {1.10 ± 0.20} & \underline{0.43 ± 0.07} & \underline{0.65 ± 0.13} & {\XSolidBrush} & {\XSolidBrush} & {\XSolidBrush}\\
FRB~20230402B & 43.6 & 3.92  & [400.2, 619.2] & \underline{0.58 ± 0.03} & \underline{0.26 ± 0.05} & {0.29 ± 0.05} & {0.37 ± 0.17} & {\XSolidBrush} & {\XSolidBrush} & {\XSolidBrush}\\
\hline
\end{tabular}}
\caption{\label{tab1} Column 1: FRB name of TNS; Column 2: total SNR of FRB; Column 3: time delays between the matched peak pairs; Column 4: frequency range for hardness test; Column 5: HR\(_{\text{ML}}\) from two episodes; Column 6: HR\(_{\text{HM}}\) from two episodes; Column 7:  optimal flux ratio; Column 8: lens redshifted mass; Column 9: microlensing candidate rank.
Hardness test results for the 9 FRBs, a deviation exceeding the $1\sigma$ region is marked in underline. Only two candidates pass the hardness test $k_{\rm HR}\geq 1$, i.e., FRB~20190131D and FRB~20211115A.}
\end{table*}

The detailed screening procedure is as follows: Firstly, we perform an ACF analysis on the light curve of each FRB dynamic spectrum. When a spike exceeding the $3\sigma$ threshold is identified, we search among the $n_{\rm peaks}$ brightest peaks in the light curve for a peak pair whose time delay matches the spike within an error tolerance of $2~{\rm ms}$. It is worth noting that if a spike is present but no corresponding peak pair can be found among these brightest $n_{\rm peaks}$ peaks, we exclude the lensing possibility. It is because the remaining peaks have very low SNR and are not considered meaningful for next analysis. If such peak pair is identified, we impose the following criteria: (i) the main peak must precede the secondary peak in time; (ii) the ${\rm PSNR}$ of secondary peak must be greater than 10 ($\mathrm{PSNR}_2 > 10$) to ensure sufficient strength for subsequent analyses; and (iii) all the peak pairs must include the global maximum ${\rm PSNR}$ peak among all detected peaks. For any peak pair that satisfies these three initial criteria, we further examine whether significant frequency drift exists between the two peaks. Peak pairs that exhibit no significant frequency drift are tentatively regarded as lensed candidates for further investigation. FRB~20181028A is excluded under the above criteria because its ACF spike is dominated by peaks 2 and 3, and the matched pair lacks the global PSNR maximum as shown in Fig.~\ref{fig3}. Based on the above criteria, i.e., ACF spikes, the arrival time order, and frequency drift, we initially screened 340 multi-peak FRBs for possible lensed candidates, identifying 11 candidates~\footnote{All of the above and subsequent analysis procedures and results are published on Zenodo:~\dataset[doi:10.5281/zenodo.22021707]{https://doi.org/10.5281/zenodo.22021707}~\citep{zhou_2026_22021707}.}. However, upon reexamining these 11 FRBs, we found that in several cases the peaks were not clearly separated, appearing more like a single integrated signal with the peaks manifesting as spiky substructures (FRB~20221129B, FRB~20221216A), or exhibiting excessively diffuse frequency profiles for corresponding peak pair (FRB~20221129B). Consequently, under an almost rigorous initial screening, ultimately only 9 FRBs were selected for subsequent analysis: FRB~20190131D, FRB~20190915E, FRB~20200603B, FRB~20210117D, FRB~20210130C, FRB~20211115A, FRB~20220225C, FRB~20220424C, FRB~20230402B.

\subsubsection{Hardness Test}
\begin{figure*}
    \centering
    \includegraphics[width=0.45\textwidth, height=0.36\textwidth]
    {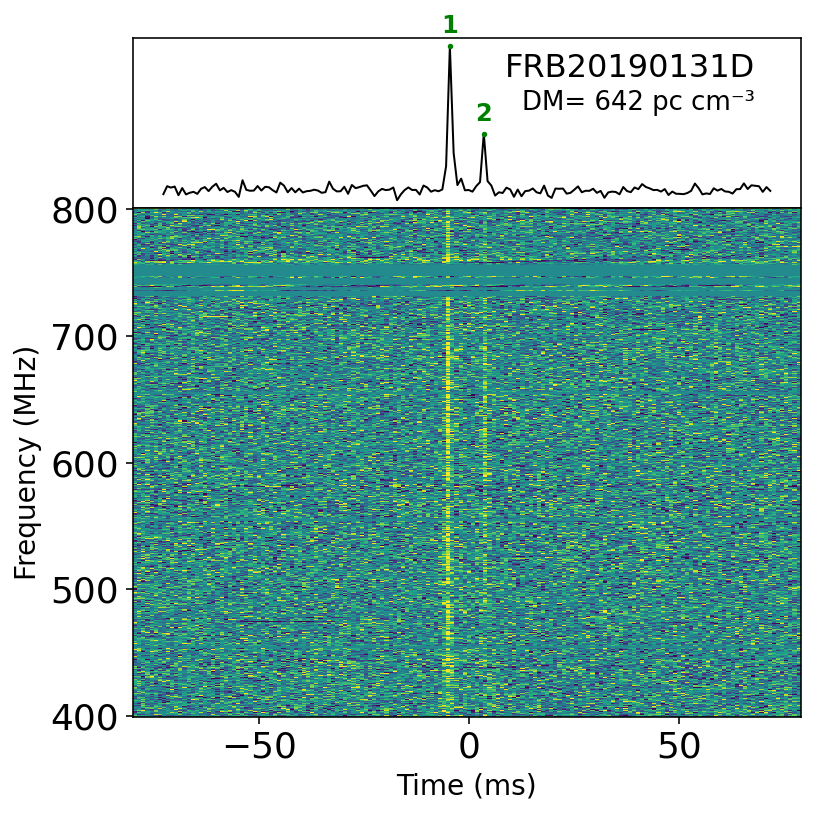}
    \includegraphics[width=0.45\textwidth, height=0.36\textwidth]
    {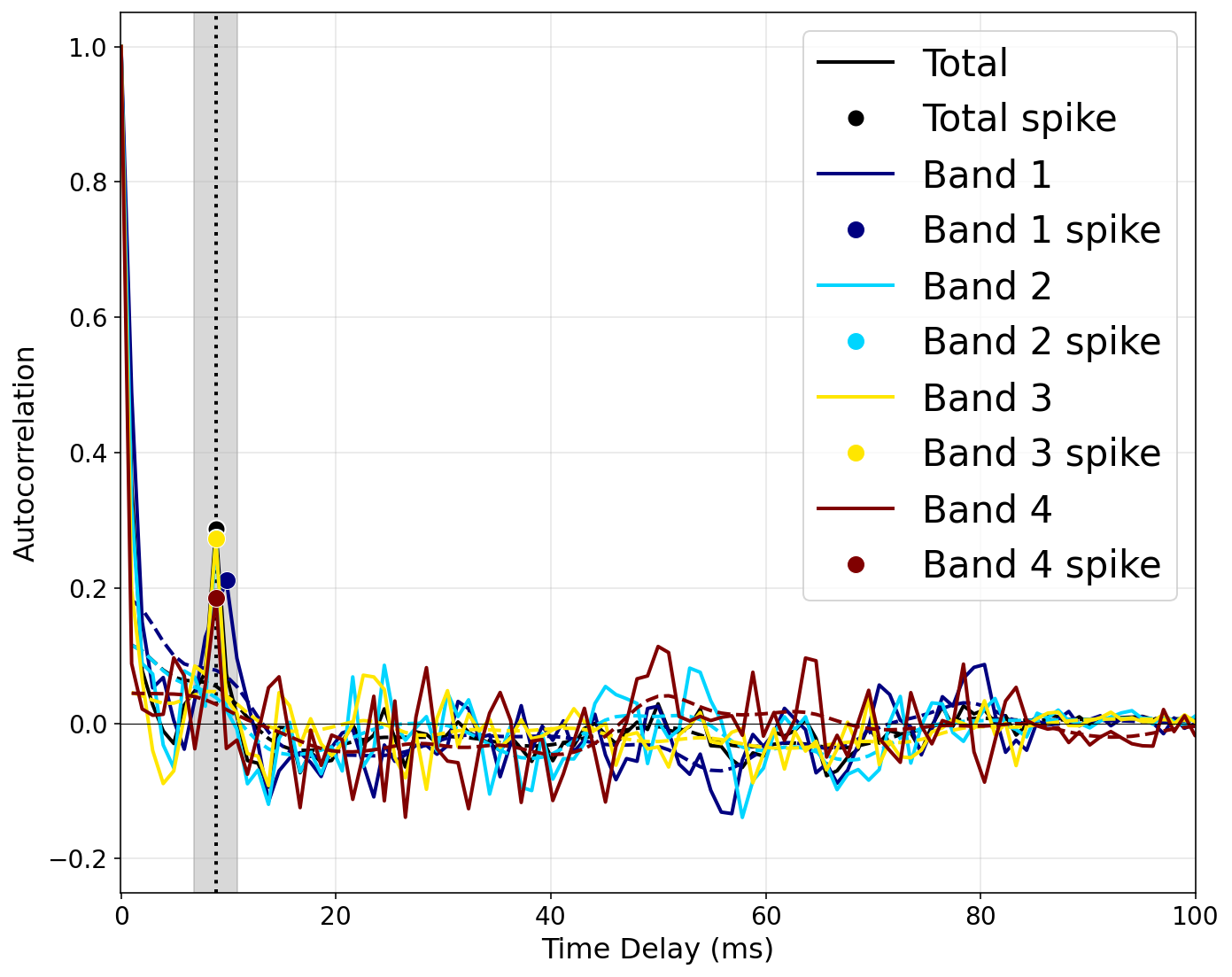}
    \includegraphics[width=0.45\textwidth, height=0.36\textwidth]
    {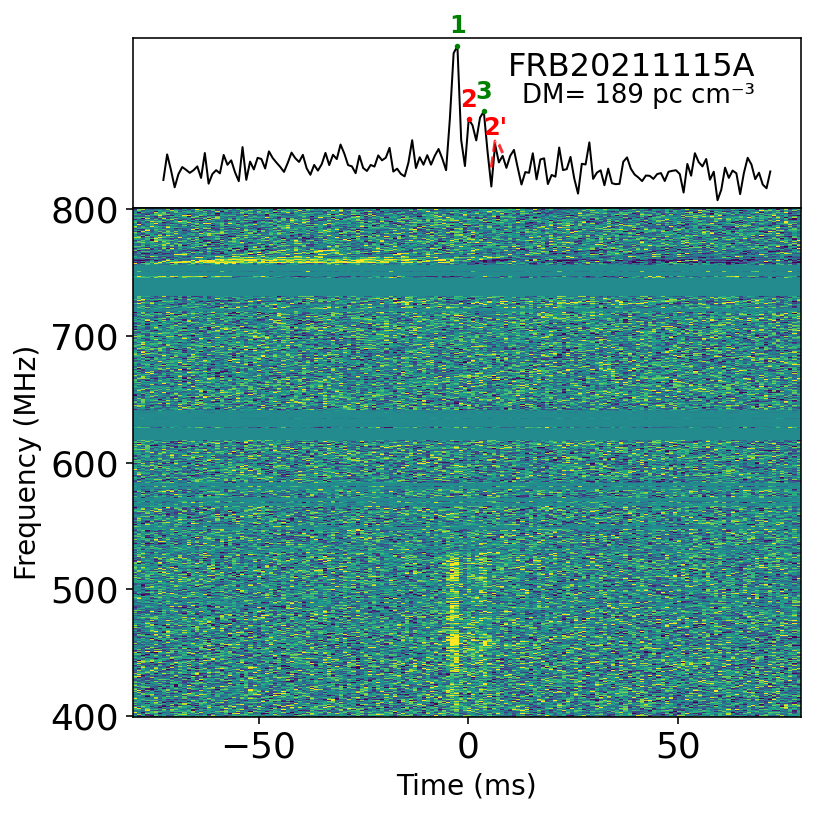}
    \includegraphics[width=0.45\textwidth, height=0.36\textwidth]
    {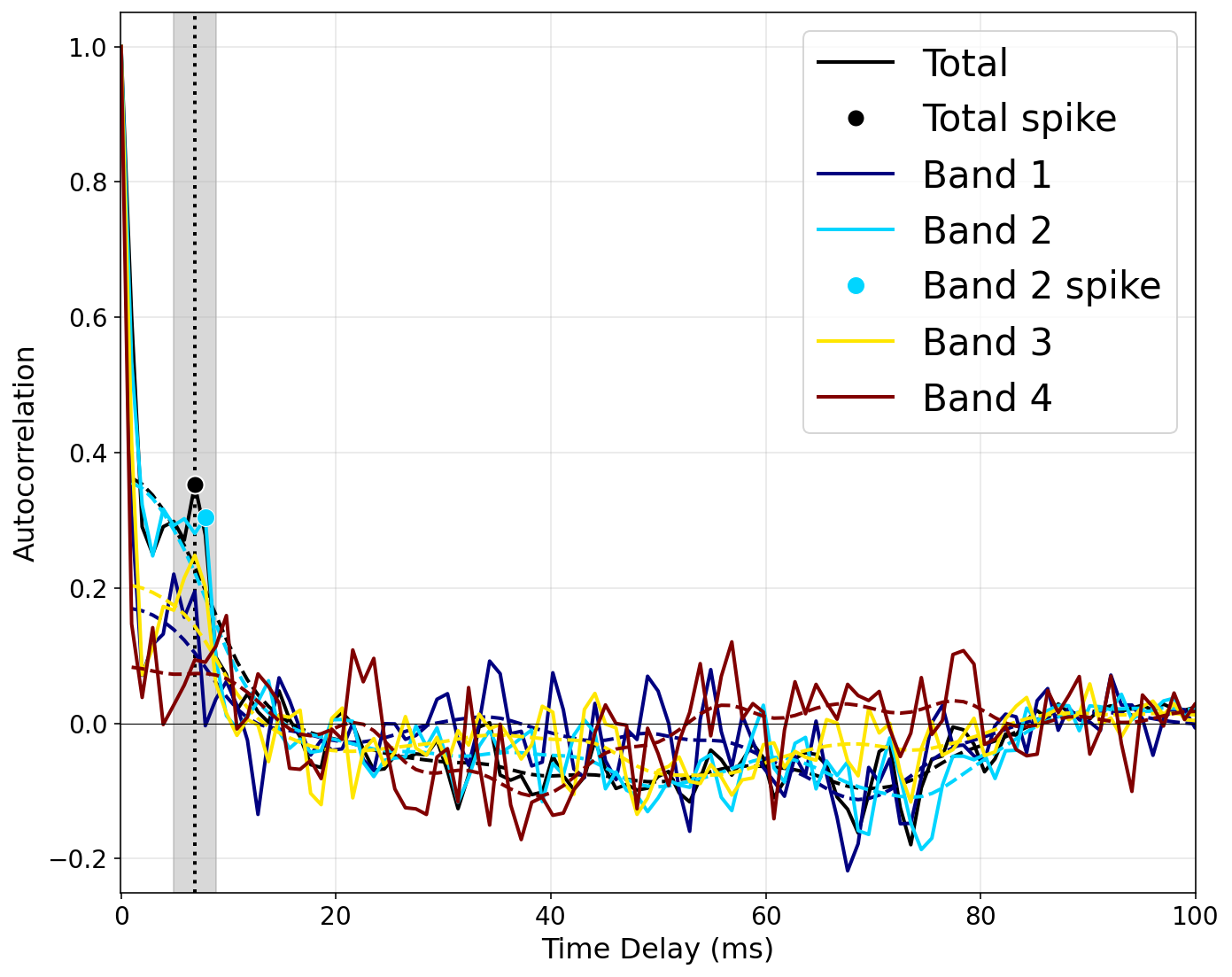}
     \caption{Similar to Upper panel of Fig.~\ref{fig3}, but for FRB~20190131D, FRB~20211115A. In the lower-left dynamic spectrum for FRB~20211115A, peak~2 is shifted to the dashed peak based on the inferred flux ratio and time delay, and is labeled as $2'$. The right panel shows the ACF with spikes within $\pm2~\rm ms$ range for the total and 4 frequency bands independently, where a colored dot on a spike indicates that the ACF exceeds the $3\sigma$ smooth threshold.
     }\label{fig4}
\end{figure*}

The hardness test has been widely used to justify the lensing effect~\citep{Paynter:2021wmb,Veres:2021gfr,Wang:2021ens,Yang:2021wwd,Lin:2021hae}, based on the hypothesis that the flux ratio between gravitationally lensed pulses should be frequency independent~\citep{Paczynski1987}. In this work, following the approach commonly adopted for lensed gamma-ray bursts (GRBs), we rebin the frequency channels into $k_{\rm HR}+2$ broad bands $(k_{\rm HR}\in\{1,2,3,\dots\})$. When \(k_{\rm HR}=1\), these bands are denoted by L, M, and H, corresponding to low, medium, and high frequencies, respectively. This rebinning procedure is designed for the hardness test: the frequency range of each FRB signal is determined from the RFI-free portion of the spectrum, i.e., $f\in\big[f_{\mathrm{low}}, f_{\mathrm{high}}\big]$, as provided in the CHIME/FRB Catalog 2. For each preliminary candidate, we first extract the time‑resolved flux profiles in different frequency channels. Based on the autocorrelation analysis of the total burst profile, we split each frequency‑resolved light curve into two similar episodes. We define a second episode of the same duration as the first one but shifted by a certain time delay. Subsequently, for each episode we define hardness ratios as
\begin{equation}\label{eq2-8}
{\rm HR}_{ij}=\frac{I_{i}-B_{i}}{I_{j}-B_{j}}~(j=i-1,~i\in [2,\dots, k_{\rm HR}+2]),
\end{equation}
where $I_{i}$ and $B_{i}$ represent the total intensity (or flux) of the burst light curve and the background noise level for one episode, respectively. The uncertainty of ${\rm HR}_{ij}$ could be estimated considering Gaussian noise and the error propagation formula as follows
\begin{equation}\label{eq2-9}
\sigma_{\text{HR}_{ij}} = \mathrm{HR}_{ij}\sqrt{ \frac{\sigma_{I_i}^2 + \sigma_{B_i}^2}{(I_i - B_i)^2} + \frac{\sigma_{I_j}^2 + \sigma_{B_j}^2}{(I_j - B_j)^2} }.
\end{equation}
Here we require the preliminary candidate to pass the hardness test only when their ${\rm HR}_{ij}$ for different episodes are consistent with the mean value within the $1\sigma$ region. To determine the signal regions for the hardness test, we combined three times the full width at half maximum (FWHM) of the main peak with the time-delay positions. For FRB~20211115A, since a substructure (peak~2) is located between peak~1 and peak~3, we adopted only one FWHM instead of three. In the hardness test, we set the threshold at $k_{\rm HR,min}=1$ (i.e., we require $k_{\rm HR} > 1$) for the low, medium, and high-frequency bands. The final analysis results, presented in Table~\ref{tab1} and Fig.~\ref{fig4}, yield two potential non-repeating FRBs that pass the hardness test, i.e., FRB~20190131D and FRB~20211115A. For both events, the maximum $k_{\rm HR,max}$ value reaches 2, which corresponds to the 4 frequency bands adopted in the hardness test.

\subsection{Analysis of FRB Microlensing Candidates}\label{sec2-3}
\subsubsection{FRB~20190131D}
For FRB~20190131D, the dynamic spectrum is shown in the upper-left panel of Fig.~\ref{fig4} with the primary and secondary peaks labeled by green numbers. At the maximum value of $k_{\rm HR,max}=2$, the ${\rm HR}_{ij}$ values in the two episodes are consistent within $1\sigma$. In addition, the ACFs for the 4 frequency bands all exceed the $3\sigma$ threshold at the same time delay as the total light curve, yielding $k_{\rm ACF}=4$ shown in the upper-right panel of Fig.~\ref{fig4}. Based on these results, we define a composite rank for microlensed FRB candidates as
\begin{equation}\label{eq2-9f}
S_{\rm Lens}\equiv k_{\rm HR,max} + k_{\rm ACF},
\end{equation}
which reflects the overall microlensing evidence and allows a direct comparison among candidates summarized in Table~\ref{tab1}. If we relax the hardness test criterion by accepting any candidate whose ${\rm HR}_{ij}$ values are consistent within $2\sigma$ (For FRB~20190915E and FRB~20220424C, the ${\rm HR}_{\rm HM}$ values are consistent within the $2\sigma$ range), then the maximum $k_{\rm HR,max}$ reaches 7 for FRB~20190131D. The signal can be divided into 9 frequency bins, while only 6 of them pass the ACF test at a significance level exceeding $3\sigma$ ($S_{\rm Lens}=7+6=13$). As the number of frequency bins increases, the SNR per bin inevitably decreases (see the total SNR values in Table~\ref{tab1}), and beyond a certain division the noise contribution becomes dominant, which limits both the hardness test and the ACF test. It is noteworthy that~\citet{Zhou:2021ndx} also analyzed this source in detail but ultimately excluded it. Unlike our present approach which adopts a coarse four-bands division, their two-dimensional cross-correlation analysis divided the signal into 512 frequency channels, resulting in large errors and hence only weak correlation.

By adopting ${\rm DM_{host}}+{\rm DM_{src}} \in[0,200]$ according to Eq.~(\ref{eq2-1}), the redshift range of the source is inferred to be $z_{\rm s}\in[0.51,0.67]$. The lens mass $M_{{\rm L},z}$ is estimated from the following expression based on the point mass model
\begin{equation}\label{eq2-10}
M_{{\rm L},z}=\frac{\Delta t}{2\bigg(\frac{R_{\rm f}-1}{\sqrt{R_{\rm f}}}+\ln R_{\rm f}\bigg)},
\end{equation}
where $R_{\rm f}$ is the ampliﬁcation ratio between two images, which could be calculated with of the burst light curve and the background noise of two episodes ($R_{\rm f}=(I_{1}-B_{1})/(I_{2}-B_{2})$). Given the lens redshift mass and the inferred redshift range, the lens mass range is estimated to be $M_{\rm L}\in[280, 467]~M_{\odot}$ falls in the IMBH regime.

\subsubsection{FRB~20211115A}
For FRB~20211115A, the situation presents different and more intriguing features. As shown in the dynamic spectrum in the lower-left panel of Fig.~\ref{fig4}, peaks 1 and 3 form a primary–secondary peak pair that satisfies the lensing signal, while peak 2 can be regarded as a substructure of peak 1. Shifting peak 2 to its corresponding image position (peak $2'$) according to the flux ratio and time delay given in Table~\ref{tab1} places it at the red dashed line, where the light curve appears to exhibit a faint and ambiguous counterpart signal. However, owing to effect of noise, the spectral signature of the image corresponding to peak 2 is not clearly resolved in the two-dimensional dynamic spectrum. As shown in the lower right panel of Fig.~\ref{fig4}, among the ACFs for the 4 frequency bands, only band 2 shows a spike at the same time delay as the total light curve above the $3\sigma$ threshold (while 3 spikes would be found if the threshold were reduced to $2\sigma$). The microlensing candidate rank for this event is $S_{\rm Lens}=2+1=3$. Similar to FRB~20190131D, if we relax the hardness test criterion by accepting candidates whose ${\rm HR}_{ij}$ values are consistent within $2\sigma$, then for this source the maximum $k_{\rm HR,max}$ reaches 4. The signal can be divided into 6 frequency bins, of which only 2 bands pass the ACF test at a significance level exceeding $3\sigma$, giving a microlensing candidate rank $S_{\rm Lens}=4+2=6$. Therefore, compared with FRB~20190131D, FRB~20211115A not only shows a similar trend of decreasing test validity with finer band divisions, but also yields lower lensing significance, which may be explained by its relatively lower SNR. The redshift range of FRB~20211115A and and lens mass ranges are obtained as $z_{\rm s}\in[0,0.13]$ and $M_{\rm L}\in[539,609]~M_{\odot}$, respectively. Similarly to the case of FRB~20190131D, this lens mass again falls within the IMBH range.

\section{Constraints on PBHs}\label{sec3}
The angular separation between different lensed images is of order $10^{-6}\,^{\prime\prime}$ for stellar-mass lenses, $1\,^{\prime\prime}$ for galaxy lenses, and $1\,^{\prime}$ for galaxy cluster lenses. The differences in path length and gravitational potential lead to time delays between these images of order $10^{-5}$~seconds for stellar-mass lenses, months for galaxy-scale lenses, and years for galaxy-cluster lenses. In~\citet{Munoz:2016tmg}, it was first pointed out that the microlensing FRBs can be used to probe the compact dark matter, e.g. PBHs with masses as small as $\mathcal{O}(10~M_{\odot})$. The PBH can be treated as a point mass with an Einstein radius $\theta_{\rm E}$. Although the spatial resolution in radio observations can reach a high level of $\sim 10^{-2}\,^{\prime\prime}$, it is still insufficient to distinguish split images for stellar-mass lenses. However, one can directly measure the time delay between lensed FRB signals. The formula for the time delay $\Delta t$ is determined by the lens redshifted mass $M_{{\rm L},z}=(1+z_{\rm L})M_{\rm PBH}$, and the impact parameter $y=\beta/\theta_{\rm E}$
\begin{equation}\label{eq3-1}
\begin{split}
\Delta t=4M_{{\rm L},z}\bigg[\frac{y}{2}\sqrt{y^2+4}+\ln\bigg(\frac{\sqrt{y^2+4}+y}{\sqrt{y^2+4}-y}\bigg)\bigg].
\end{split}
\end{equation}
$\Delta t$ must be larger than the width ($w$) of the observed signal. This requires $y$ to be larger than a certain value $y_{\rm min}(M_{\rm PBH},z_{\rm L},w)$ according to Eq.~(\ref{eq3-1}). The lensing cross section due to a PBH lens is then given by:
\begin{equation}\label{eq3-2}
\begin{split}
\sigma(M_{\rm PBH}, z_{\rm L}, z_{\rm S}, w)=\frac{4\pi M_{\rm PBH}D_{\rm L}D_{\rm LS}}{D_{\rm S}}\times\\
\left[y^2_{\rm max}(R_{\rm f,max})-y^2_{\rm min}(M_{\rm PBH},z_{\rm L},w)\right].
\end{split}
\end{equation}
The maximum value of the normalized impact parameter is found by requiring that the flux ratio of the two lensed images exceeds a reference value $R_{\rm f,max}$
\begin{equation}\label{eq3-3}
y_{\rm max}(R_{\rm f,max})=R_{\rm f,max}^{1/4}-R_{\rm f,max}^{-1/4}.
\end{equation}
$R_{\rm f,max}$ should depend on the SNR of each observed FRB, which is reasonable since both SNR and flux ratio are crucial for identifying a lensed FRB event. The lensing optical depth due to a single PBH is:
\begin{equation}\label{eq3-4}
\begin{split}
\tau(M_{\rm PBH},f_{\rm PBH},z_{\rm S},w)=\frac{3}{2}f_{\rm PBH}\Omega_{\rm DM,0}\int_0^{z_{\rm S}}dz_{\rm L}\frac{H_0^2}{H(z_{\rm L})}\times\\
\frac{D_{\rm L}D_{\rm LS}}{D_{\rm S}}
(1+z_{\rm L})^2[y^2_{\rm max}(R_{\rm f,max})-y^2_{\rm min}(M_{\rm PBH},z_{\rm L},w)],
\end{split}
\end{equation}
where $H(z_{\rm L})$ is the Hubble function, $f_{\rm PBH}$ represents the fraction of PBHs in dark matter, and $\Omega_{\rm DM,0}$ is the present density parameter of dark matter. Given the statistically meaningful total FRB number $N_{\rm tot}$, the expected number of lensed FRBs can be approximated by the sum of the lensing optical depths of all FRBs.
\begin{equation}\label{eq3-5}
N_{\rm lensed~FRB}=\sum_{i=1}^{N_{\rm tot}}\tau_i(M_{\rm PBH},f_{\rm PBH},z_{{\rm S},i},w_i).
\end{equation}

\begin{figure*}
    \centering
    \includegraphics[width=0.75\textwidth, height=0.45\textwidth]
    {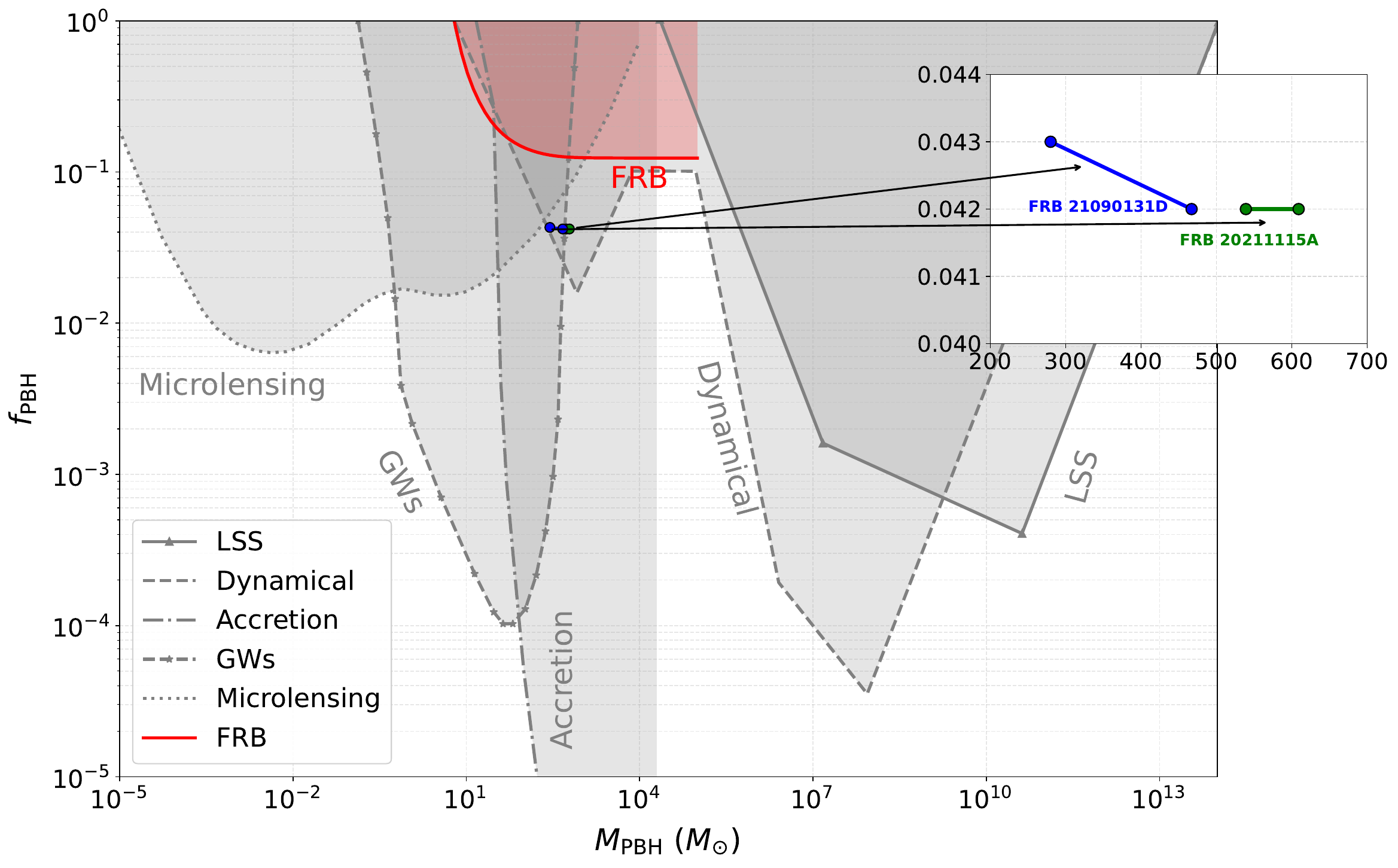}
     \caption{The upper limit on the PBH fraction $f_{\rm PBH}$ at the $95\%$ confidence level as a function of PBH mass $M_{\rm PBH}$ in the range $[1,10^{5}]\,M_{\odot}$, derived from the current CHIME/FRB catalog 2 under the assumption of no lensing FRBs. The subplot presents an approximate constraint on the PBH parameters under the hypothesis that FRB~20190131D and FRB~20211115A are lensed signals produced by PBHs. Other constraints are compiled from existing review~\cite{2026NCimRtmp4C}: these include limits from microlensing surveys (Microlensing), merger rate of binary black holes (GWs), effect of accretion (Accretion), dynamical effects (Dynamical), and the imprint of PBHs on large-scale structure (LSS).
     }\label{fig5}
\end{figure*}

As shown in Fig~\ref{fig1}, the width distribution and the SNR distribution ($R_{\rm max,f}={\rm SNR}/10$) together determine the magnitude of the scattering cross section in Eq.~(\ref{eq3-2}). In addition, we find that the inferred redshifts of FRBs tend to be concentrated at low redshifts ($z<2$) in Fig~\ref{fig1}, which is one of the factors that significantly influences the results. If no genuine microlensed signal is found in the current data, the curve in the $(M_{\rm PBH}, f_{\rm PBH})$ parameter space that predicts at least three detectable lensed signals should be ruled out at the $95\%$ confidence level. As shown in Fig~\ref{fig5}, the mass can be tested down to $\sim10~M_{\odot}$, and $f_{\rm PBH}$ is gradually constrained to $13\%$ for the intermediate-mass PBH range $\gtrsim300~M_{\odot}$ at the $95\%$ confidence level by using SNR-dependent flux ratio thresholds. Owing to the increase in the number of CHIME/FRB sources from $\sim5\times10^2$ to $\sim4\times10^3$, our result constitutes an improvement of nearly an order of magnitude over previous constraints derived from CHIME/FRB catalog 1~\citep{Zhou:2021ndx,Krochek:2021opq}. Although current constraints are relatively weak, especially for small masses, much tighter constraints will be obtained from a large number of high-SNR FRBs in the near future~\citep{Munoz:2016tmg,Laha:2018zav,Connor:2022bwl}. Assuming both FRB~20190131D and FRB~20211115A originate from intermediate-mass PBH lensing system, a simple estimate of the PBH abundance in the relevant mass ranges can be made. We find that the PBH abundances in the mass ranges $\sim[280,467]~M_{\odot}$ and $\sim[539,609]~M_{\odot}$ are both constrained to approximately $4\%$, which are in significant tension with other observational constraints, e.g., effect of accretion (Accretion) and dynamical effects (Dynamical) as shown in Fig.~\ref{fig5}.

\section{Conclusions and Discussions}\label{sec4}
In this paper, we develop a pipeline to identify microlensed FRBs based on their dynamic spectra and apply it to FRB data from the CHIME/FRB Catalog 2, searching for microlensing signatures. The tests reveal evidence of correlation between the peaks of these FRBs. Intriguingly, we have reported two potential microlensing candidates, i.e. FRB~20190131D and FRB~20211115A. The inferred lens masses for these two signatures are $\sim[280,467]~M_{\odot}$ and $\sim[539,609]~M_{\odot}$, respectively. Both of them are in the range of IMBH. So far, there is no literature reporting the presence of galaxies or galaxy clusters in their line of sight directions of these two candidates. It indicates that the two IMBHs might be isolated and of primordial origins. On the basis of the primordial origin, we obtained a preliminary constraint on the PBH dark matter fraction: 1) these PBHs would constitute $\sim4\%$ of the dark matter in the mass ranges $\sim[280,467]~M_{\odot}$ and $\sim[539,609]~M_{\odot}$; 2) if these candidates are not true gravitational lensing signals, the abundance of PBHs with masses $>300~M_{\odot}$ could be constrained to $\sim13\%$ at the $95\%$ confidence level. 

Notably,~\citet{Paynter:2021wmb} identified a milli-lensing signal in the light curve of GRB~950830 and inferred a lens mass of $\sim10^4~M_{\odot}$, which lies within the IMBH range. Meanwhile, the GW event GW231123\_135430 has been analyzed as a possible binary black hole lensed by a $\sim600~M_{\odot}$ lens~\citep{LIGOScientific:2025cwb}. Together with the two possible IMBH candidates in our work, and given that FRB~20190320B was confirmed as a microlensed candidate with lens mass $\sim400~M_{\odot}$ by analyzing high‑time‑resolution channelized data~\citep{Xiong:2025gtw}, the number of evidence for the existence of IMBHs is constantly increasing. Alternative interpretations should also be considered. The observed sub‑bursts could arise from intrinsic emission processes, such as double‑peaked profiles produced by magnetospheric activity or propagation effects within the FRB environment. Furthermore, as noted by~\citet{Cordes:2017eug}, plasma structures within FRB host galaxies can also refract and lensing radio signals, giving rise to time delays and interference patterns analogous to gravitational lensing, thus offering a propagation-based explanation for FRB substructures and repetitions. Therefore, in future studies, it is essential to consider not only more complex lensing systems but also additional physical mechanisms that can produce lensing-like signals. Robust models of FRB light curves and spectral functions, incorporating Bayesian analyses and spectral index analyses analogous to those in searching for lensed GRBs (Fast-Rise Exponential-Decay pulse model consistent with GRB fireball internal shocks~\citep{Paynter:2021wmb,Wang:2021ens,Lin:2021hae,Yang:2021wwd,Veres:2021gfr}), would provide additional evidence for validating microlensed FRBs. Complementary observations, such as polarization measurements or higher time resolution (a capability that suffers from a fundamental limitation of CHIME's strategy for surveying FRBs, i.e., the sparse sampling around the peak of the burst), will be crucial for distinguishing genuine lensing events from false positives and for robustly validating microlensed FRBs.

\section{Acknowledgements}
We thank the anonymous reviewer for his/her valuable comments, which helped us to improve the quality of the manuscript. This work is supported by National Key R\&D Program of China under Grant No.2024YFC2207400; Science and Technology Research Project of Hubei Provincial Department of Education under Grants No.Q20251302; Research Performance Assesssment Grant of the Postdoctoral Fellowship Program of China Postdoctoral Science Foundation under Grant No.YJB20250367; National Natural Science Foundation of China under Grants Nos.12322301, 12275021, and 12503002.

\bibliography{ref}{}
\bibliographystyle{aasjournal}

\end{document}